\documentclass{aa}  

\usepackage{graphicx}
\usepackage{txfonts}
\usepackage{graphicx}	
\usepackage{amsmath}	
\usepackage{amssymb}	
\usepackage{gensymb}
\usepackage{textgreek}
\usepackage{booktabs}
\usepackage{xcolor}
\usepackage{placeins}
\usepackage[colorlinks=true, citecolor=blue]{hyperref}

\begin{document}

   \title{The kinematic contribution to the cosmic number count dipole}

   \author{J.D. Wagenveld\inst{1}
          \and S. von Hausegger\inst{2}
          \and H-R. Kl\"{o}ckner\inst{1}
          \and D.J. Schwarz\inst{3}
          }

   \institute{Max-Planck Institut fur Radioastronomie, 
              Auf dem H\"{u}gel 69, 
              53121 Bonn, Germany
        \and Department of Physics, 
             University of Oxford, 
             Parks Road, Oxford OX1 3PU, UK
        \and Fakult\"{a}t f\"{u}r Physik, 
             Universit\"{a}t Bielefeld, 
             Postfach 100131, 
             33501 Bielefeld, Germany
             }


    \abstract{Measurements of the number count dipole with large surveys have shown amplitudes in tension with kinematic predictions based on the observed Doppler dipole of the cosmic microwave background (CMB). These observations seem to be in direct conflict with a homogeneous and isotropic Universe as asserted by the cosmological principle, demanding further investigation into the origin of the tension. Here, we investigate whether the observed number count dipoles are consistent with being fully kinematic, regardless of boost, or if there is any residual anisotropy contributing to the total observed dipole, independent of the kinematic part. To disentangle these contributions, we aim to leverage the fact that the kinematic matter dipole expected in a given galaxy catalogue scales with observed properties of the sample, and different catalogues used in the literature therefore have different kinematic dipole expectations. We here perform joint dipole fits using the NRAO VLA Sky Survey (NVSS), the Rapid ASKAP Continuum Survey (RACS), and the AGN catalogue derived from the Wide-field Infrared Survey Explorer (CatWISE). The direction of the common dipole between these catalogues is offset from the CMB dipole direction by $23\pm5$ degrees. Assuming a common kinematic and non-kinematic dipole component between all catalogues, we find that a large residual, non-kinematic dipole anisotropy is detected, though a common direction between the two components is disfavoured by model selection. Freeing up both amplitude and direction for this residual dipole while fixing the kinematic dipole to the CMB dipole expectation, we recover a significant residual dipole with $\mathcal{D}_{resid} = (0.81\pm0.14)\times10^{-2}$, that is offset from the CMB dipole direction by $39\pm8$ degrees. While these results cannot explain the origin of the unexpectedly large number count dipoles, they offer a rephrasing of the anomaly in terms of kinematic and non-kinematic contributions, providing evidence for the existence of the latter within the models explored here. The present work provides a valuable first test of this concept, although its scrutinising power is limited by the currently employed catalogues. Larger catalogues, especially in radio, will be needed to further lift the degeneracy between the kinematic and residual dipole components.} 


   \maketitle
%

\section{Introduction}

Modern cosmological models based on Friedmann-Lema\^{i}tre-Robertson-Walker (FLRW) metrics, such as \textLambda-CDM, are built on the assumptions of the cosmological principle. As such, these models require homogeneity and isotropy on the largest scales. While the cosmic microwave background (CMB) is a remarkable example of this large scale isotropy, with fluctuations around the CMB monopole being largely as small as 1 part in $10^5$, the CMB dipole appears to be an exception. A hundred times larger than the smaller-scale fluctuations, the CMB dipole is conventionally considered to be the result of our movement with respect to the frame in which the CMB would have appeared isotropic, the so-called CMB rest frame. Under this interpretation, measurements of the CMB dipole translate a velocity of the Solar System with respect to the CMB of $v = 369.82\pm0.11\ \mathrm{km\ s^{-1}}$ \citep{Aghanim2020}.

Our movement is expected to result in an apparent dipole also in the number counts of cosmologically distant sources, whose rest frame ought to agree with that of the CMB \citep{Ellis1984}. Caused by the relativistic effects of aberration, Doppler boosting, and Doppler shifting of the observed source positions and spectra, the same physics is at play as that suspected at the root of the CMB dipole. This immediately places an expectation on the kinematic matter dipole: a dipole pointing in the same direction as the CMB dipole with an amplitude proportional to $\beta=v/c$, where $c$ is the speed of light. To measure the expected $\mathcal{O}(10^{-3})$ number count dipole, it was predicted that catalogues of extragalactic sources in excess of $10^6$ were required to reach $3\sigma$ statistical significance \citep{Crawford2009}. The earliest statistically significant measurements were performed with a catalogue of radio sources from the National Radio Astronomy Observatory (NRAO) Very Large Array (VLA) Sky Survey \citep[NVSS,][]{Condon1998}, where the measured dipole consistently had a larger amplitude than expected \citep[e.g.][]{Blake2002,Singal2011,Rubart2013,Siewert2021,Secrest2022,Wagenveld2023b}, while being broadly consistent with the CMB dipole in terms of direction. Other radio catalogues have since been used for this measurement with varying levels of success, yielding similar results, though in absence of independent measurements at other wavelengths it was difficult to exclude common systematic effects as the cause of the dipole. 

An important breakthrough in the credibility of these measurements was the recent addition of a measurement of the number count dipole with infrared AGN \citep{Secrest2021,Secrest2022,Dam2023}. This measurement was performed with a catalogue of sources observed by the Wide-field Infrared Survey Explorer \citep[WISE,][]{Wright2010}. As WISE is a space-based telescope, it is not influenced by any of the potential systematic effects that would affect ground-based radio observations. Furthermore, the sample of sources was entirely independent from NVSS and other radio catalogues. This measurement yielded a dipole amplitude that was two times larger than the kinematic expectation, with a significance of $4.9\sigma$. With this level of significance, these measurements are in serious tension with the cosmological principle. Since then, additional measurements of the dipole with quasars selected in a composition of optical and infrared measurements \citep{Mittal2024,Mittal2024a} also showed a higher amplitude than expected, although to a less significant degree, due to lower source counts. With this, an excess dipole has now been confirmed at different wavelengths, with different instruments, and in different independent source samples.

With such pronounced excess dipoles measured in these catalogues, the question arises whether the measured dipole is purely caused by the velocity of the observer, as is assumed in \citet{Ellis1984}. Rather than a purely kinematic `EB dipole', some other (residual) component could be contributing to the total observed dipole, that does not scale with $\beta$. Such a question lies at the core of studies employing a range of methods to scrutinise the kinematic dipole signals detected (and suspected) thus far. For instance, the CMB Doppler dipole is expected to be accompanied by aberration of the CMB anisotropies and though indeed corresponding measurements report consistency \citep{PlanckCollaboration2014,Saha2021,Ferreira2021}, it has not been conclusively shown that the CMB dipole is entirely kinematic. For the number count dipole, methods to measure kinematic and non-kinematic components separately have been proposed \citep[e.g.][]{Nadolny2021,Tiwari2015}, although these require a great deal more data than what is currently available. Only recently, \citet{Ferreira2024} and \citet{Tiwari2024} attempted to measure the kinematic dipole directly using redshifts from Sloan Digital Sky Survey (SDSS). While there is an indication for consistency with the CMB dipole, large uncertainties remain. Separate measures of the Solar system velocity, for example with SNIa \citep[e.g.][]{Horstmann2022}, seem to favour consistency with the velocity obtained from the CMB dipole as well. However, an explicit separation of kinematic and non-kinematic dipoles has not been made, and these components have so far not been independently and simultaneously measured. 

We here propose and implement an alternative method for isolating the kinematic dipole, as expected from \citet{Ellis1984}, from a potential non-kinematic, residual dipole, by using multiple catalogues at the same time. We achieve this by utilising the dipole estimator from \citet{Wagenveld2023b}, which is able to fit a common dipole signal from a set of catalogues. This avoids the problems that stem from attempting to combine the catalogues, due to differences in frequency, angular resolution, and flux density limit \citep[e.g.][]{Colin2017, Darling2022}. Previously, this procedure allowed a combined dipole estimate between NVSS and the Rapid Australian Square Kilometre Array Pathfinder (ASKAP) Continuum Survey \citep[RACS-low,][]{Hale2021}. This measurement could be performed because the predicted (kinematic) dipole amplitude of both catalogues was the same. While true for the case of NVSS, RACS-low and other radio catalogues, this is not the case in general. For instance for CatWISE the expected dipole amplitude is nearly twice that of NVSS and RACS-low. Under certain assumptions, chief of which is that any residual non-kinematic dipole component is common between the catalogues, using catalogues with different expected kinematic dipole amplitudes allows a combined estimate to be used to separate these components.

In this paper, we will use the NVSS, RACS-low and CatWISE AGN catalogues and their different expected kinematic dipole amplitudes to isolate the kinematic component of the number count dipole. The paper is organised as follows. In Section~\ref{sec:estimators} we describe the estimators, and in Section~\ref{sec:data} we describe the data we will use to perform this measurement. The results are presented in Section~\ref{sec:results}, and discussed in Section~\ref{sec:discussion}. We conclude in Section~\ref{sec:conclusion}.

\section{The cosmic number count dipole}
\label{sec:estimators}

As a result of Doppler boost, Doppler shift and relativistic aberration induced by our velocity with respect to background sources, we see a dipole in the observed number counts of extragalactic background sources. The amplitude of this kinematic number count dipole, caused by the velocity of the observer $\beta=v/c$, is given by \citep{Ellis1984}
\begin{equation}
    \mathcal{D}_{kin} = [2 + x(1+\alpha)]\beta,
    \label{eq:dipole_amplitude}
\end{equation}
and depends on the spectral index of the sources, $\alpha$\footnote{Here we use the spectral index convention $S\propto\nu^{-\alpha}$.}, and the power law index of the flux density distribution, $x$. These parameters in general differ per catalogue and survey, and as such the expected kinematic dipole amplitude for a given $\beta$ can also differ. Given the fact that the measured dipole amplitude has in many cases been larger than the expected kinematic dipole amplitude, the possibility arises that the excess dipole amplitude is not kinematic, but is caused by a different phenomenon altogether. In this case (assuming that both components point in the same direction), we can hypothesise that the total dipole amplitude can be broken down as
\begin{equation}
    \mathcal{D} = \mathcal{D}_{kin} + \mathcal{D}_{resid},
    \label{eq:dipole_components}
\end{equation}
where $\mathcal{D}_{kin}$ is the \citet{Ellis1984} prediction from Equation \ref{eq:dipole_amplitude}, and $\mathcal{D}_{resid}$ represents the contribution of a residual, non-kinematic dipole component. While the fact that the total dipole we measure is relatively close to the CMB dipole in direction indicates that any residual dipole might also point in a similar direction, it is unlikely that the separate components do point in the exact same direction. We distinguish corresponding considerations following Equation~\ref{eq:dipole_components_big} below.

\subsection{Dipole estimation}

For dipole estimation we expand upon the existing Bayesian estimators introduced in \citet{Wagenveld2023b}. These estimators are based on the fact that the counts-in-cells distribution of sources isotropically distributed across the sky follows a Poisson distribution
\begin{equation}
    p(n) = \frac{\lambda^ne^{-\lambda}}{n!},
\end{equation}
where $n$ is the number of sources in a cell, and $\lambda$ represents the mean and variance of the distribution. The most basic Poisson estimator assumes only a dipole and monopole, which affect $\lambda$ as
\begin{equation}
    \lambda(\mathcal{M},\vec{d}) = \mathcal{M}(1+\vec{d}\cdot\vec{\hat{n}}).
\end{equation}
Here $\mathcal{M}$ represents the monopole, and $\vec{d}$ the dipole vector, the amplitude of which is $\mathcal{D}$. In order to estimate the dipole parameters, we maximise the likelihood given by
\begin{equation}
    \mathcal{L}(\vec{n}|\vec{d},\mathcal{M}) = \prod_i \frac{\lambda(\mathcal{M},\vec{d})^{n_i}e^{-\lambda(\mathcal{M},\vec{d})}}{n_i!},
\end{equation}
over all cells $i$. This likelihood was shown in \citet{Wagenveld2023b} to produce similar results as quadratic estimators, and outperforms them in the limit of low counts where the assumption of Gaussian noise no longer holds. It was also shown that this basic estimator can be extended in several ways to increase the number of usable sources for a dipole measurement while accounting for systematics. One such extension of the basic Poisson estimator was used in \citet{Wagenveld2024a} to fit for an additional linear relation between a specific parameter $y$ and the source density
\begin{equation}
    \lambda(\vec{d}, \mathcal{M}, \varepsilon, y) = \mathcal{M}[1 - \varepsilon\cdot y](1+\vec{d}\cdot\hat{n}),
    \label{eq:poisson_linear}
\end{equation}
where $\varepsilon$ is defined as the slope of the linear relation. Below, we use this estimator to fit the observed change in source density as a function of absolute ecliptic latitude seen in CatWISE \citep{Secrest2021}, similar to what was done in \citet{Dam2023}. As demonstrated in \citet{Dam2023}, not including the ecliptic latitude effect is heavily disfavoured by model selection, and yields an even higher dipole amplitude than if it is included. 

\subsection{Joint dipole estimation}

Another extension of the basic Poisson estimator is the multi-Poisson estimator, which can take multiple catalogues and perform a joint dipole estimate. This estimator assumes a common dipole but a different monopole for each catalogue, thereby allowing the fit of a common dipole signal. The likelihood of this estimator is defined as
\begin{equation}
    \mathcal{L}(\vec{n}|\vec{d},\vec{\mathcal{M}}) = \prod_j \left[\prod_i \frac{\lambda(\mathcal{M}_j,\vec{d})^{n_{i,j}}e^{-\lambda(\mathcal{M}_j,\vec{d})}}{n_{i,j}!}\right],
    \label{eq:multi_poisson}
\end{equation}
taking the product over each cell $i$ in each catalogue $j$ and taking the product of all catalogues. The efficacy of this estimator is however predicated on the fact that the catalogues have the same dipole signal, whereas the amplitude of the kinematic dipole actually depends on the spectral indices and flux density distribution of the sources in the catalogue. As such, if we wish to combine catalogues with different expected kinematic dipole amplitudes, this presents an opportunity to, under certain assumptions, actually isolate the kinematic dipole from any other components contributing to the dipole, following Equation~\ref{eq:dipole_components}. Based on this principle, we redefine the dipole vector $\vec{d}$ in Equation~\ref{eq:multi_poisson} as
\begin{equation}
    \vec{d}_j = [2 + x_j(1+\alpha_j)]\vec{\beta} + \vec{d}_{resid},
    \label{eq:dipole_components_big}
\end{equation}
for catalogue $j$, where each catalogue has a different pair of $x$ and $\alpha$ values. To get a better understanding of what model describes the data the best, we consider several different hypotheses. 
\begin{enumerate}[(i)]
    \item[(0)] The dipole consists of a single component, which is fully kinematic.
    \item The dipole consists of two components, one kinematic and one residual, which both point in the same direction. 
    \item The dipole consists of two components, one kinematic and one residual, which both point in different directions.
    \item The dipole consists of two components, one kinematic and one residual, which both point different directions. The kinematic component is fixed to what is expected from the CMB dipole in both amplitude and direction.
\end{enumerate}
Here, hypothesis (0) is considered our ``null'' hypothesis, and seeks to explain the observed dipole purely by kinematics. In all other cases, we separate the kinematic contribution and residual dipole, described by $\vec{\beta}$ and $\vec{d}_{resid}$, respectively. If the direction of these components is the same, Equation~\ref{eq:dipole_components_big} simply becomes a linear equation with two unknowns, and thus requires at least two measurements with significantly different outcomes to resolve. This is of course more complicated once we associate uncertainties with these measurements. In the latter two cases, we also allow for separate directions for these two components. We stress here that the separation of dipole components as presented here assumes that different catalogues have the same $\vec{d}_{resid}$, even though the origin of this dipole component is not known. The validity of this assumption is discussed in Section~\ref{sec:discussion}.

\subsection{Priors}

To avoid biasing results, we aim to make priors as uninformed as possible. For the dipole, we separately fit the right ascension and declination of the dipole direction, as well as the dipole amplitude. We do not restrict or fix any particular direction (unless explicitly mentioned), so the priors on right ascension and declination of the dipole direction are the same for both the kinematic and residual dipole. On individual catalogues we can only fit for the total dipole amplitude $\mathcal{D}$, while in the multi-Poisson estimator, we fit for both $\beta$ and $\mathcal{D}_{resid}$. We define these priors such that the total dipole amplitude can not exceed unity. Regardless of estimator, we estimate the monopole $\mathcal{M}$ for each catalogue separately, using as an initial estimate the mean of all cell counts $\bar{n}$. We can summarise these general priors as follows
\begin{align*}
    \pi(\mathcal{D}) &\sim u \\
    \pi(\mathcal{D}_{resid}) &\sim 0.5\cdot u \\
    \pi(\beta) &\sim 0.05\cdot u \\
    \pi(\mathrm{R.A.}) &\sim 360\cdot u \\
    \pi(\mathrm{Dec.}) &\sim \sin^{-1}[2u-1] \\
    \pi(\mathcal{M}) &\sim 2\bar{n}\cdot u. 
\end{align*}
Here, $u = \mathcal{U}[0,1]$ represents a uniform distribution between 0 and 1. For the Poisson estimator implementing a linear fit described in Equation~\ref{eq:poisson_linear}, we fit for an additional parameter $\varepsilon$. The only requirement for this parameter is that the resulting number counts should not be negative, making the prior dependent on the maximum value of $y$, such that $\pi(\varepsilon) \sim (2u-1)/y_{max}$.

All above described esimators are implemented using the Bayesian inference library \textsc{bilby} \citep{Ashton2019}. Through \textsc{bilby}, we maximise the likelihood with MCMC sampling using \textsc{emcee} \citep{Foreman-Mackey2013}. After sampling, the best-fit parameters are obtained by taking the median of the posterior distribution, with the uncertainties represented by the 16\% (lower) and 84\% (upper) quantiles of the distribution. The scripts where these have been implemented are available on GitHub\footnote{\url{https://github.com/JonahDW/Bayesian-dipole}} and an immutable copy is archived in Zenodo \citep{Wagenveld2025}. 
For the purposes of model comparison, we use \textsc{harmonic}, which implements the learnt harmonic mean estimator \citep{McEwen2021} to compute the marginal likelihood, $\mathcal{Z}$.

\section{Data}
\label{sec:data}

\begin{table*}
    \renewcommand*{\arraystretch}{1.4}
    \centering
    \caption{Best fit dipole estimates on the individual NVSS, RACS, and CatWISE catalogues.}
    \begin{tabular}{c c c c c c c c}
    \hline \hline
    Catalogue & $S_0$ & $N$ & $\mathcal{M}$ & $\varepsilon$ & $\mathcal{D}$ & R.A. & Dec.\\
     & (mJy) & & counts/pixel & ($\times10^{-4}$) & ($\times10^{-2}$) & (deg) & (deg) \\
    \hline
    NVSS & 15 & 352,862 & $10.11\pm0.02$ & -- & $1.39\pm0.29$ & $151\pm12$ & $-9\pm14$ \\
    NVSS$^{\mathrm{a}}$ & 15 & 351,483 & $10.07\pm0.02$ & -- & $1.20\pm0.29$ & $152\pm14$ & $-10\pm16$ \\
    RACS-low & 15 & 442,046 & $14.20\pm0.02$ & -- & $1.43\pm0.24$ & $190\pm10$ & $3\pm13$ \\
    RACS-low$^{\mathrm{a}}$ & 15 & 440,377 & $14.15\pm0.02$ & -- & $1.34\pm0.24$ & $194\pm11$ & $8\pm14$ \\
    CatWISE & 0.078 & 1,567,586 & $68.23\pm0.09$ & $9.2\pm0.4$ & $1.51\pm0.16$ & $141\pm5$ & $-6\pm6$ \\ \hline
    \end{tabular}
    \tablefoot{$^{\mathrm{a}}$ Excluded sources matched to the 2MRS catalogue.} 
    \label{tab:results}
\end{table*}

For the purposes of testing these hypotheses we want to combine several catalogues that have different expected kinematic dipole amplitudes. These catalogues should also provide reliable dipole measurements on their own, and thus be large enough to yield a significant dipole measurement individually. Furthermore, for combined measurements, we want to select catalogues that can easily be made statistically independent from each other. In a purely kinematic interpretation of the number count dipole, the dipole signal is dominated by sources near the flux density limit. As such, covering an completely independent sample of sources is not necessary for a statistically independent measurement. If however an intrinsic dipole is thought to contribute to the overall dipole, then this signal can originate from the overall observed source population. This motivates the use of the RACS-low, NVSS and CatWISE catalogues, which have all independently yielded robust and significant measurements of the number count dipole. With NVSS and RACS-low covering the northern and southern hemisphere radio populations respectively, and CatWISE covering the infrared quasar population, these catalogues see for the most part different sources. As such, they are easily made completely statistically independent by removal of shared sources, enabling joint dipole measurements to constrain the contribution of a residual dipole effect. 

For all catalogues, in order to predict the kinematic dipole amplitude, we must determine the index of the power law, $x$, describing the integral flux density distribution of sources, and the (average) spectral index of sources, $\alpha$. Because the dipole effect is dominated by sources near the lower flux density limit, these values should be measured in the vicinity of this limit \citep{vonHausegger2024}. In all cases, these values are determined after the final data cuts described in Section~\ref{sec:mask}. We note here that as we determine the values here, they are not included as free parameters in the dipole estimation, as they would add a significant number of additional nuisance parameters. As such, though these values can be noisy, their variance is not included in the modeling, which can in principle lead to an underestimation of the overall uncertainty in the other (dipole) parameters. However, we expect this variance to be highly suppressed given the fact that they are measured for the entire catalogue \citep[as seen in][]{Secrest2022}.  

\subsection{NVSS}

The NVSS catalogue \citep{Condon1998} was the first catalogue on which a significant dipole measurement could be performed \citep{Blake2002}, and was also the first catalogue for which an excess dipole amplitude was claimed \citep{Singal2011}. It covers the entire sky north of a declination of $-40\degree$, at a frequency of 1.4~GHz and an angular resolution of $45\arcsec$. The full catalogue, which includes the Galactic plane, contains 1,773,484 sources. To homogenise the catalogue for a dipole measurement we make a flux density cut at 15~mJy. The flux density distribution around this flux density cut is well fit by a power law with an index $x=0.74$. While no spectral indices are available for NVSS, at a frequency of 1.4~GHz the mean spectral index is commonly found to be around 0.75 \citep[e.g.][]{Condon1984}. As such, we adopt $\alpha=0.75$ for NVSS, which sets the expectation for a purely kinematic dipole with a canonical velocity of $\beta=1.23\times10^{-3}$ to $\mathcal{D}=0.41\times10^{-2}$.

\subsection{RACS-low}

RACS \citep{McConnell2020} covers the sky south of a declination of $+30\degree$, and is being released in several stages at different frequencies. The first of these data releases is RACS-low, which was first used for a dipole measurement in conjunction with the VLA Sky Survey \citep[VLASS,][]{Lacy2020} by \citep{Darling2022}, however the first dipole measurement with only RACS-low was performed by \citet{Wagenveld2023b}. RACS-low has an angular resolution of $25\arcsec$, and a central frequency of 887.5~MHz. The RACS-low catalogue, which does not include the Galactic plane, contains 2,123,638 sources. To homogenise the catalogue for a dipole measurement we make the same flux density cut as for NVSS, at 15~mJy. The flux density distribution around this flux density cut is well fit by a power law with an index $x=0.72$. Flux density comparisons between RACS-low, NVSS, and RACS-mid indicate steeper spectral indices for these sources \citep{Hale2021,Duchesne2023}, as the lower frequency observations will detect more sources with steeper spectra. Thus, for RACS-low, we adopt $\alpha=0.88$ \citep[the median $\alpha$ found in][]{Duchesne2023}, which translates to a kinematic dipole amplitude of $\mathcal{D}=0.41\times10^{-2}$ for our canonical velocity, identical to the expectation for NVSS.  

\begin{figure*}[t]
    \centering
    \sidecaption
    \includegraphics[width=0.7\textwidth]{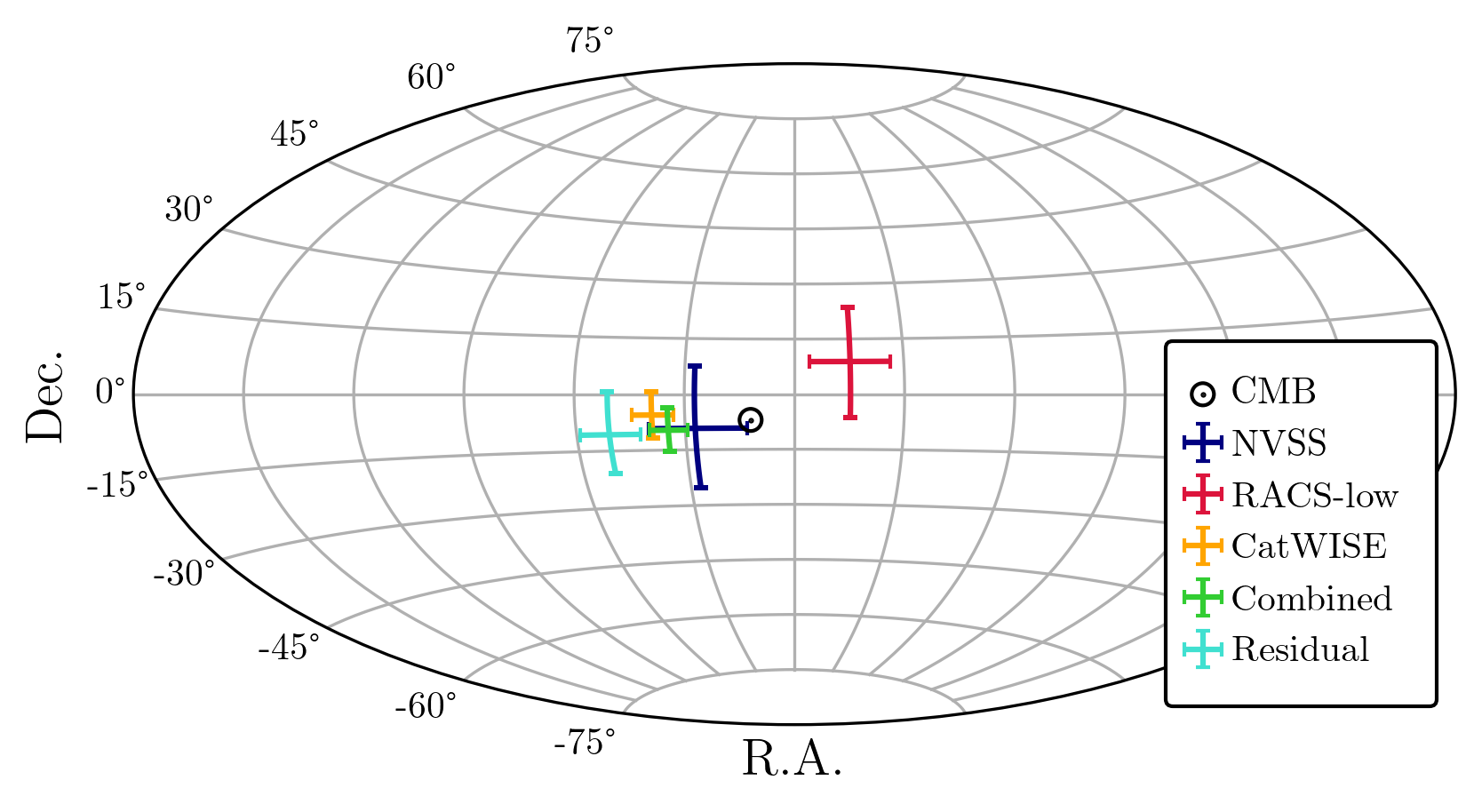}
    \caption{Best fit dipole directions for the individual catalogues NVSS (blue), RACS-low (red), and CatWISE (orange). For the combined fits, the best fit direction for the full dipole (hypothesis (0) and (i), green) and the residual dipole fixing the kinematic dipole to the CMB dipole (hypothesis (iii), turquoise) are also shown. In all cases, the errorbars indicate the 1$\sigma$ uncertainties. The direction of the CMB dipole is indicated by the black circle and dot.}
    \label{fig:dipole_directions}
\end{figure*}

\subsection{CatWISE}

WISE \citep{Wright2010} is an infrared space-based telescope that has observed the whole sky. For a dipole measurement at infrared wavelengths, extragalactic sources must be reliably selected from the observations, which also contain stars and dusty galaxies. The CatWISE2020 catalogue, which contains more than a billion sources, is the most recent and largest catalogue compiled from these observations \citep{Marocco2021}. The AGN selected from this catalogue were used for the dipole measurement in \citet{Secrest2021}, with updated version of the catalogue, which we will use here, being used for the dipole measurement in \citet{Secrest2022}. In this present catalogue, AGN are selected with a $W1 - W2 \geq 0.8$ cut, which leaves 4,145,046 sources in the catalogue, including the Galactic plane. To homogenise the catalogue a cut of $W1 < 16.5$ was made, which corresponds to a flux density cut of 0.078~mJy. The flux density distribution of this WISE catalogue around this flux density cut is well fit by a power law with an index $x=1.90$. Spectral indices are available for all the CatWISE sources, however they do not follow a Gaussian distribution. Furthermore, spectral indices are generally lower for sources with lower flux densities. For sources below 0.08~mJy, the mean spectral index is $\alpha=1.07$, so that is the value we use for the WISE catalogue. Here, a purely kinematic dipole with the canonical velocity would have an amplitude of $\mathcal{D}=0.73\times10^{-2}$.

\subsection{Masking and pixelisation}
\label{sec:mask}

Though generally most sources in these catalogues are expected to be part of the extragalactic background, there is a possibility of nearby sources to contaminate the signal. \citet{Oayda2024a} showed that excluding sources in the NVSS and RACS-low catalogues that coincided local sources in the 2MASS Redshift Survey \citep[2MRS,][]{Huchra2012} catalogue had an appreciable impact on the recovered dipole amplitude, indicating that local structure might positively contribute to the observed dipole amplitude \citep[for earlier studies see also][]{Colin2017,Rameez2018}. As this is undesirable, we choose to exclude these sources as well. We do this by matching the sources of our selected catalogues with all sources at $z<0.02$ in the 2MRS catalogue, using the angular resolution of each survey as a matching radius. This removes 4681 sources from NVSS, 4314 from RACS, and 66 from CatWISE. This is done before any flux density cuts. We choose to apply a redshift limit of $z=0.02$, as beyond this the redshift distribution of 2MRS drops off significantly, indicating that it is no longer volume-complete. Furthermore, as the \citet{Ellis1984} test assumes a redshift integrated source distribution, redshift cuts lead to boundary effects that modify the expected dipole amplitude \citep{vonHausegger2024a}. Since in this work we rely on accurate dipole amplitude expectations, these boundary terms are especially relevant here. While their accurate prediction requires good knowledge of the redshift distributions of these catalogues, we can estimate their impact for our specific case to be small.\footnote{We can estimate that, for a catalogue like NVSS, the removal of sources with redshift of $z < 0.02$ leads to a reduction of the expected dipole amplitude of $\mathcal{O}(1\%)$, whereas $z < 0.1$ the reduction is $\mathcal{O}(10\%)$.}.

After applying the aforementioned flux density cuts for all of the catalogues, there are still undesired sources or areas with anomalous source counts. For all catalogues, sources associated with the Milky Way should be excluded, which is accomplished by masking low Galactic latitudes. For NVSS and RACS-low we exclude $|b| < 7\degree$, and following \citet{Secrest2022} we exclude $|b| < 30\degree$ in the CatWISE catalogue. While this is a significant area that is masked for CatWISE, it can be shown that Galactic latitude masks as large as $\pm30\degree$ do not produce any biased dipole estimates for source densities as large as those considered here \citep[see e.g.][]{Oayda2025}. Additional masked areas for NVSS and RACS-low are the same as in \citet{Wagenveld2023b}, and for these catalogues sources within a radius of $0.3\degree$ from any source brighter than 2.5 Jy are masked. For CatWISE the masked areas detailed in \citet{Secrest2022} are also masked here. 

Following masking, the catalogues are pixelised using the Hierarchical Equal Area isoLatitude Pixelization \citep[\textsc{HEALPix},][]{Gorski2005} scheme, which divides the sky into equal sized cells. The resolution of the sky map is determined by the $N_{side}$ parameter, with the number of cells in the whole sky being $N_{pix}=12\times N_{side}^2$. Here we choose a resolution of $N_{side}=64$, for which the cells are $55\arcmin$ on a side. The value assigned to each cell is the number of sources contained within it. We mask all cells with zero sources in it, as well as all cells bordering a cell with zero sources to remove edge effects.

\begin{figure*}
    \centering
    \includegraphics[width=0.65\textwidth]{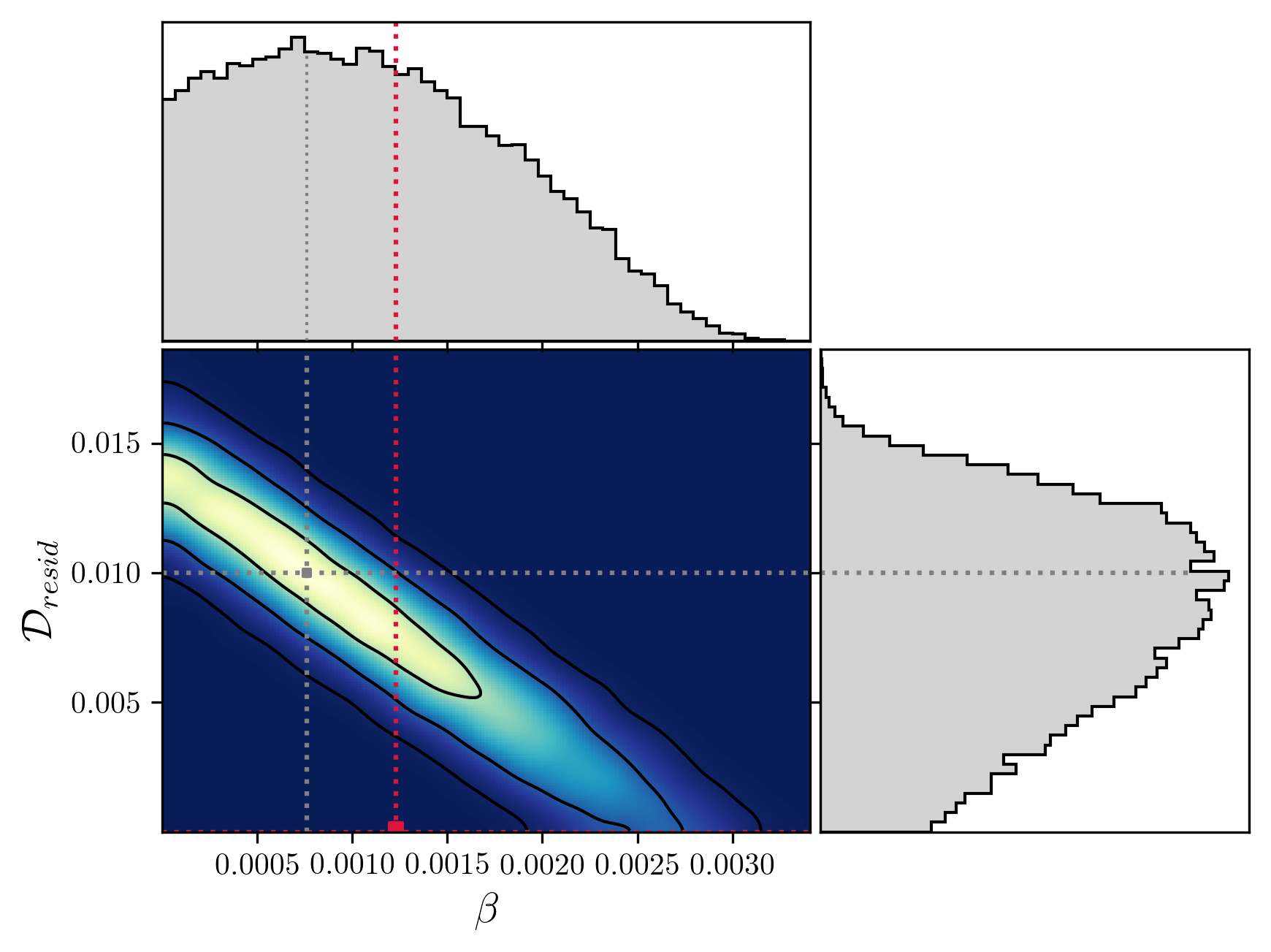}
    \includegraphics[width=0.33\textwidth]{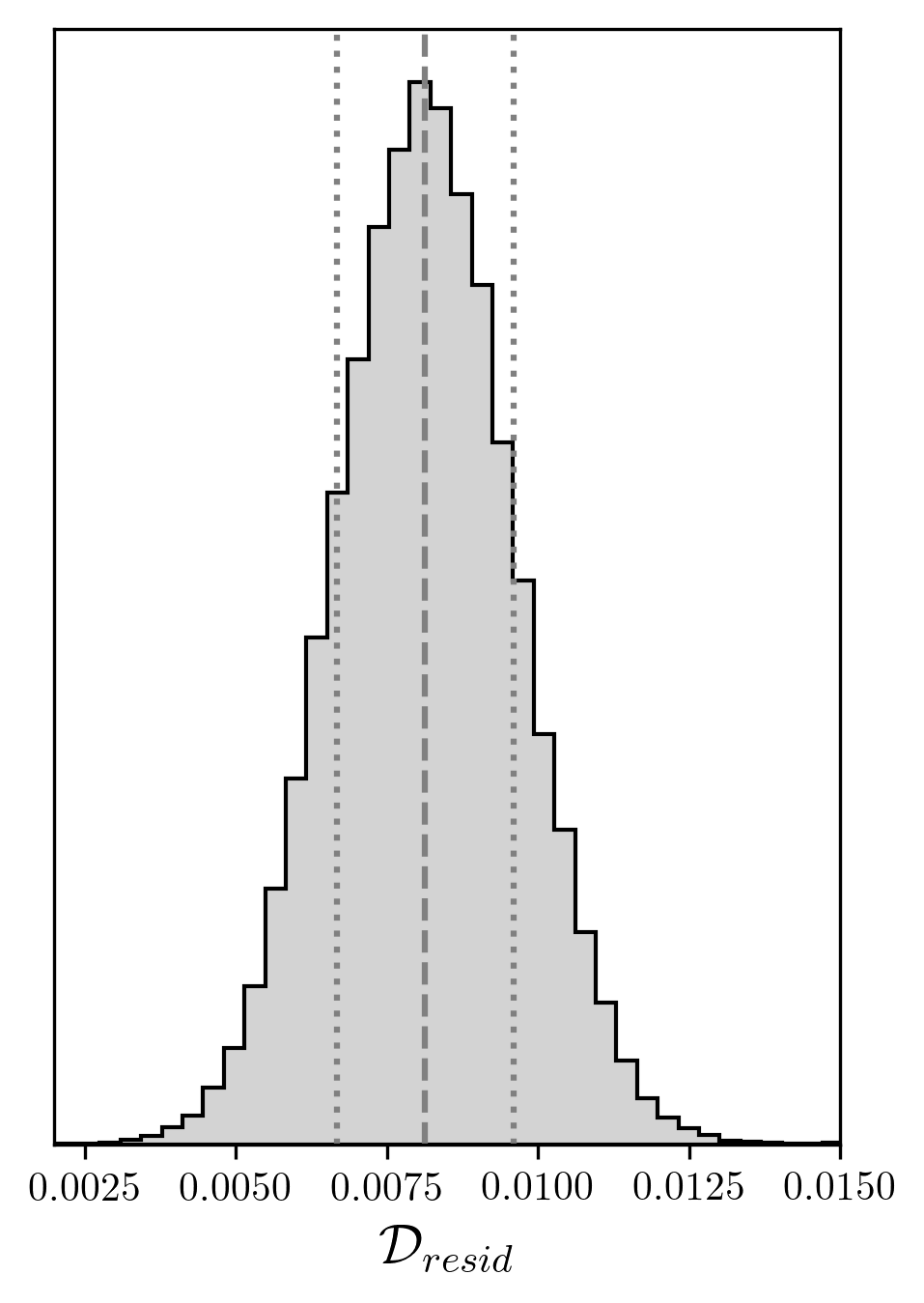}
    \caption{Left: Hypothesis (i). Posterior distribution of $\beta$ and $\mathcal{D}_{resid}$ from the combined estimate using NVSS, RACS-low, and CatWISE. The 1-,2-, and 3-$\sigma$ uncertainties are indicated by the black contours. The dotted lines indicate the maximum posterior values for these parameters. The canonical CMB velocity of $\beta=1.23\times10^{-3}$ is indicated by the dotted red line, the red dot indicating the kinematic dipole expected in the standard cosmology, assuming a negligible structure dipole ($\mathcal{D}_{resid}\approx0$). Right: Hypothesis (iii). Posterior distribution of $\mathcal{D}_{resid}$, fixing the kinematic dipole to the CMB dipole. Note the respective evidences for and against these posteriors' hypotheses in Table 2.}
    \label{fig:beta_dipole}
\end{figure*}

\section{Results}
\label{sec:results}

Using the estimators described in Section~\ref{sec:estimators}, we perform dipole estimates on the NVSS, RACS-low, and CatWISE catalogues. The resulting best fit dipole parameters for the individual catalogues are shown in Table~\ref{tab:results}. Where 2MRS sources have not been removed, as the masking and pixelisation are identical for RACS-low and NVSS, the resulting dipole parameters match those from \citet{Wagenveld2023b}. With 2MRS sources excluded, we see that for both NVSS and RACS-low the dipole amplitude is reduced, as was seen in \citet{Oayda2024a}. Since CatWISE only matched 66 sources with 2MRS, there is no appreciable change in the recovered dipole parameters. For CatWISE, there is a known linear trend in source counts as a function of absolute ecliptic latitude \citep{Secrest2021}, so we include a linear fit for this trend in the parameter estimation, following \citet{Dam2023}. For later combined fits, we create a map of weights following the best-fit linear relation for the ecliptic latitude effect (shown in Table~\ref{tab:results}). Since this effect is independent of the other catalogues, this allows us to account for it without having to fit for it separately in each hypothesis.

Figure~\ref{fig:dipole_directions} shows the resulting individual best fit dipole directions, showing that the dipole direction of the CatWISE catalogue closely matches that of NVSS. Though the dipole result of RACS-low is roughly consistent with both other catalogues in terms of amplitude, it is offset from the NVSS and CatWISE dipole by $\Delta\theta\sim 45\degree$. Nevertheless, both the NVSS and RACS-low dipole directions match that of the CMB dipole within $3\sigma$. Only the CatWISE catalogue yields small enough uncertainties to be significantly offset from the CMB dipole.

\subsection{Combined dipole estimates}

For a dipole estimate combining NVSS, RACS-low, and CatWISE, the same masks and flux density cuts are used as the estimates on the individual catalogues shown in Table~\ref{tab:results}. In addition however, to ensure that these catalogues are statistically independent, we exclude all NVSS sources in the southern hemisphere and exclude all RACS-low sources in the northern hemisphere, such that these catalogues have no overlap. We also cross-match both NVSS and RACS-low with the CatWISE catalogue, and remove sources that match within the angular resolution of the respective catalogues from the CatWISE catalogue. With this, we have three completely independent catalogues with which we can perform joint dipole fits. Combining these three catalogues yields 2,035,541 sources, the largest sample used for a dipole estimate thus far. We use the multi-Poisson estimator described in Section~\ref{sec:estimators} to obtain best-fit dipole parameters, using Equation~\ref{eq:dipole_components_big}. We again emphasize here that our hypotheses are all based on the assumption that the only kinematic contribution to the combined dipole stems from Equation~\ref{eq:dipole_amplitude}, i.e. no other dipole contributions that scale with $\beta$ are considered. We discuss the implication of this assumption in Section~\ref{sec:discussion}.

The designated null hypothesis assumes that the observed dipole is entirely kinematic. This assumption yields a best-fit dipole direction shown by the green cross in Figure~\ref{fig:dipole_directions}. Since CatWISE is contributing more than half of the sources, the direction is closely aligned with the dipole direction of the CatWISE catalogue. The direction is offset from the CMB dipole direction by $23\pm5$ degrees, with a significance of $4.6\sigma$. As shown in Table~\ref{tab:combined_results}, the best-fit value for the velocity is $\beta=(2.62\pm0.26)\times10^{-3}$ or $v=786\pm78$ km /s, which is a factor of 2.1 larger than the canonical velocity obtained from the CMB dipole. Comparison with the results of the individual catalogues shows that this closely aligns with the possible boost of the CatWISE catalogue, but differs from the boosts of the NVSS and RACS-low catalogues. Of course, these differences could (to an extent) be due to shot noise. To determine to which extent these latter catalogues disfavour a purely kinematic explanation, we must consider the hypotheses which fit for a non-kinematic component. As our null hypothesis, the marginal likelihood from this fit serves as a baseline against which to compare the other models. Thus, for each subsequent hypothesis we also present in Table~\ref{tab:combined_results} the Bayes factor, defined as the difference between the natural logarithm of the marginal likelihood of that hypothesis and this null hypothesis, $\ln B = \Delta \ln \mathcal{Z}$.

\begin{table*}
    \renewcommand*{\arraystretch}{1.4}
    \centering
    \caption{Summary of results and Bayes factors for the different tested hypotheses. Since the fit did not converge for hypothesis (ii), it is not included.}
    \begin{tabular}{l c c c c c c c}
       \hline \hline
       Hypothesis & $\beta$ & R.A.$_{\beta}$ & Dec.$_{\beta}$ & $\mathcal{D}_{resid}$ & R.A.$_{\mathcal{D}_{resid}}$ & Dec.$_{\mathcal{D}_{resid}}$ & $\ln B$ \\
                  & ($\times10^{-3}$) & (deg) & (deg) & ($\times10^{-2}$) & (deg) & (deg) & \\
       \hline 
       (0)  & $2.62\pm0.26$ & $144\pm5$ & $-10\pm6$ & $0$ & -- & -- & -- \\
       (i)  & $<1.67$ & $145\pm5$ & $-10\pm6$ & $1.00\pm0.47$ & $145\pm5$ & $-10\pm6$ & -3.0 \\
       (iii) & $1.23$ & $168$ & $-7$ & $0.81\pm0.14$ & $129\pm8$ & $-11\pm11$ & 3.5 \\
       \hline
    \end{tabular}
    \tablefoot{In hypothesis (i) the directions of both components are by design of the model the same. $^{\mathrm{a}}$ Here the upper limit is computed at 84\% CL to match the CLs of the other measurements in this table.}
    \label{tab:combined_results}
\end{table*}

\subsection{Separating dipole components}

Fitting for hypothesis (i), the best-fit dipole direction is the same as before, shown by the green cross in Figure~\ref{fig:dipole_directions}. Now in addition to $\beta$, the amplitude of a residual, non-kinematic dipole component is also included in the model, given by $\mathcal{D}_{resid}$. The left panel of Figure~\ref{fig:beta_dipole} shows the posterior distribution of $\beta$ and $\mathcal{D}_{resid}$ for this estimate. There is a clear degeneracy between $\beta$ and $\mathcal{D}_{resid}$, which is to be expected given that both parameters contribute to the overall dipole. Nevertheless, there is a region of higher likelihood that allows us to constrain the preferred values of $\beta$ and $\mathcal{D}_{resid}$, as given in Table~\ref{tab:combined_results}. The distribution of $\beta$ is consistent with zero, with $\beta < 2.34\times10^{-3}$ at 95\% confidence level (CL). This is however also consistent with the canonical velocity from the CMB ($\beta=1.23\times10^{-3}$, indicated by the red line in Figure~\ref{fig:beta_dipole}). At face value, the posterior thus indicates a preference for the kinematic contribution to be consistent with or lower than the CMB velocity, rather than with a value of $\beta\sim2.6\times10^{-3}$ at which the common dipole is fully kinematic and $\mathcal{D}_{resid}\sim0$. Comparing this model to hypothesis (0) quantitatively requires us to account for the additional free parameter. The Bayes factor of $\ln B_{i,0} = -3.0$ indicates preference of the fully kinematic model over this one. Nevertheless, the model itself shows a preference for a non-zero residual dipole, which motivates us to further investigate the possibility of this component being present.

For some additional checks, we perform the same hypothesis (i) fits for a number of different variations on the datasets, the results of which are shown in Appendix~\ref{app:additional_results}. Figure~\ref{fig:beta_dipole_app1} shows the results for combining each of the radio catalogues separately with CatWISE. Since we do not have to account for overlap between NVSS and RACS-low, we do not restrict each to their respective hemisphere here. These distributions generally favour higher values of $\beta$, with peak posteriors closer to the canonical CMB velocity. As such, the fully kinematic dipole is less strongly disfavoured while the fully residual dipole is less favoured. The difference with the result shown in Figure~\ref{fig:beta_dipole} is likely caused by disagreement between the observed dipole directions of NVSS and RACS-low. Though they are expected to have very similar dipole amplitudes, the difference in the RACS-low and NVSS dipole directions causes the combination of them to have an overall lower dipole amplitude \citep[as was also shown in][]{Wagenveld2023b,Oayda2024a}, which also lowers the inferred value of $\beta$ as we see in Figure~\ref{fig:beta_dipole} (left panel). In Figure~\ref{fig:beta_dipole_app2}, we again consider all catalogues, but apply different flux density cuts to the data. The flux limits chosen here are optimised to avoid being affected by systematics in the catalogue, while still maintaining a sufficient number of sources for the dipole estimate. Nonetheless, it is useful to see the behaviour of these fits for different flux cuts. The upper panel shows the result if we increase the flux cut of all catalogues, to 20 mJy for NVSS and RACS-low, and 0.09 mJy (W1 < 16.4) for CatWISE \citep[matching][]{Secrest2021}.  With this cut, either through a genuine effect or because of shot noise induced by the decreased source density, both NVSS and RACS-low significantly increase in dipole amplitude, resulting in a lower estimate of $\beta$. This rather stark difference from our other results is surprising, and highlights the need for deeper radio catalogues. With the current data however, it is difficult to say what causes this difference. For completeness, the lower panel of Figure~\ref{fig:beta_dipole_app2} shows the results if sources matched to 2MRS sources at $z < 0.1$ are removed from all catalogues, though as mentioned previously, 2MRS is volume-incomplete at redshifts beyond $z\gtrsim0.02$. With the caveats that likely not all $z < 0.1$ sources are removed, and that boundary effects are not accounted for (as discussed in Section~\ref{sec:mask}), the dipole amplitudes of NVSS and RACS-low are lowered, resulting in higher values of $\beta$ being preferred in this fit.

If the dipole consists of different components, as Figure~\ref{fig:beta_dipole} suggests, forcing these components to point in the same direction is potentially too constraining a model. Though generally it has been found that the directions of the overall dipole are consistent with the CMB dipole direction, CatWISE, which has the most discriminating power in a single catalogue, has a dipole direction which is inconsistent with the CMB dipole. Looking at the different dipole directions in Figure~\ref{fig:dipole_directions}, we see that there is a good agreement between the CatWISE and NVSS dipole directions, indicating that if the NVSS measurement were more precise it might yield this result as well. The RACS-low dipole however is offset in a different direction, which makes the case less clear. Because NVSS and RACS-low do not agree in terms of direction, while NVSS and CatWISE agree, this provides little constraints on the directions of the separate dipole components. Freeing up both directions and amplitudes as in hypothesis (ii) thus does not yield a converging estimate. Instead, we reduce hypothesis (ii) to hypothesis (iii), fixing the kinematic dipole both in terms of direction as well as amplitude to the CMB expectation, while leaving the residual dipole completely free.

The resulting direction for the residual dipole is shown in turquoise in Figure~\ref{fig:dipole_directions}, and has an angular separation from the kinematic dipole of $39\pm8$ degrees, with a significance of $4.6\sigma$. To understand whether this offset could be due to a chance alignment, we can consider two interpretations of alignment and correlation between the dipole component directions. Firstly, we can simply consider the probability of finding two directions on a sphere with a distance of $\Delta\theta$. For this case the distance between the different directions can be between $0\degree$ (alignment) and $90\degree$ (anti-alignment), resulting in the probability $p = \frac{1}{2}[1-\cos(\Delta\theta)]$. For the above result, the probability of the two directions being within $\Delta\theta=40\degree$ of each other is $p=0.12$. This however does not consider anti-alignment as equally `special' as alignment, even though anti-alignment can be considered a correlation between the directions to the same degree that an alignment can. For this case, the probability is instead $p = 1-\cos(\Delta\theta)$, doubling the probability for our result to $p=0.24$. Thus, with the current results, there is no conclusive evidence as to whether or not the directions of the two dipole components are correlated. However, if $\mathcal{D}_{resid}$ signifies a true density fluctuation on the largest scales, it would not be hard to imagine that this density gradient could (partially) pull all matter along it, in which case an alignment between the kinematic and non-kinematic dipole component is entirely natural. For the amplitude of the residual dipole, the posterior distribution is shown in the right panel of Figure~\ref{fig:beta_dipole}. As also shown in Table~\ref{tab:combined_results}, this indicates a best-fit $\mathcal{D}_{resid}=(0.81\pm0.14)\times10^{-2}$, corresponding to a detection significance of $5.4\sigma$. Not surprisingly, this amplitude is consistent with the residual dipole common to CatWISE AGN and NVSS reported by \citet{Secrest2022}. However, the present work also computes the corresponding posterior. Comparing this model to hypothesis (0), we obtain a Bayes factor of $\ln B_{iii,0} = 3.5$, indicating this model to be favoured in comparison. This shows that separating the kinematic and non-kinematic component both in terms of amplitude and direction is favourable, though as mentioned before, converging fits can only be achieved if one component is fixed, which limits the utility of this approach until larger catalogues have been surveyed. 

\section{Discussion}
\label{sec:discussion}

The cosmic dipole anomaly, namely the unexpectedly large dipole amplitudes found in number counts of galaxies, raises the question about the origin and composition of the observed dipoles. In this work we were specifically interested in distinguishing kinematic from non-kinematic contributions, or as written here $\mathcal{D}_{kin}$ and $\mathcal{D}_{resid}$, respectively. To this effect we assumed the \citet{Ellis1984} dipole amplitude (Equation~\ref{eq:dipole_amplitude}) to be the only kinematic contribution to the dipole amplitude observed in a given galaxy catalogue; its proportionality to the continuous parameter $\beta$ allowed us to consider a fit to the observer velocity. Crucial to this work was that the factor preceding $\beta$ being fixed by the respective galaxy catalogue’s properties. Simultaneously we assumed the residual dipole amplitude to be the same in each of the employed galaxy catalogues. In combination, these assumptions help to break the degeneracy otherwise present between $\beta$ and $\mathcal{D}_{resid}$, cf. Equation~\ref{eq:dipole_components} (or alternatively Equation~\ref{eq:dipole_components_big}, which allows individual directions of the constituent dipoles).

The idea of a common residual dipole in different catalogues was initially explored in \citet{Secrest2022} who simply compared the dipoles found in the NVSS and CatWISE samples before and after subtraction of the expected kinematic dipole. Assuming the canonical observer velocity taken from the CMB dipole, yielded nearly matching values of $\mathcal{D}_{resid}$ for both catalogues, sparking our hypothesis (iii). Our work essentially formalises this comparison, accounting for uncertainties in amplitudes and directions consistently. By including the RACS-low catalogue, we further extended this ansatz to more data, using the additional sky coverage and accompanying scrutinising power. And yet, with this compilation of data, neither a purely kinematic, nor a purely residual dipole can be conclusively excluded, if the kinematic component is indeed entirely determined by Equation~\ref{eq:dipole_amplitude}. Nevertheless, we find that the most preferred model is one where the kinematic component of the number count dipole is close to or lower than the purely kinematic interpretation of the CMB dipole, i.e. $\beta\sim1.23\times10^{-3}$. The remainder of the number count dipole is correspondingly made up by the residual, non-kinematic dipole.

The general idea to separate kinematic from non-kinematic matter dipole contributions is not unique to this work. A number of observables has been considered for this purpose in recent years. For example, weighted number counts can be constructed in a way to separate an intrinsic ``clustering’’ component from that expected due to the observer boost alone \citep{Nadolny2021}; these are yet to be applied to suitable data with redshift estimates. But even utilising redshift measurements alone, without collection of number counts, can provide a handle on the observer’s velocity as per the induced Doppler shifts on the redshifts. This was recently applied to data from the SDSS by \citet{Ferreira2024} and in an attempt to reproduce their results also by \citet{Tiwari2024}. While some subsamples deliver measurements of $\beta$ in agreement with the value derived from the CMB dipole, other subsamples do not provide similar constraints, and their methods do not yet appear to match the subtleties of the data. Lastly, the angular aberration induced by the observer boost that leads to the factor of 2 in Equation~\ref{eq:dipole_amplitude} might also be considered in an isolated manner, when studying source counts. This has been worked out theoretically by \citet{Lacasa2024} with good prospects for upcoming surveys to constrain $\beta$ and therewith our velocity with respect to the matter rest frame.

\subsection{Assumptions on the present work}

It is valuable to highlight what might violate the assumptions made for this analysis.  Firstly, contributions to the expected dipole amplitudes that scale with $\beta$ in addition to those of Equation~\ref{eq:dipole_amplitude} would undermine the premise of our analysis. To date, no such contributions are known for galaxy catalogues as those investigated here. While it has been suggested recently, that redshift evolution of the catalogues’ source populations might lead to effects beyond Equation~\ref{eq:dipole_amplitude} and also proportional to $\beta$ \citep[e.g.][]{Dalang2022,Guandalin2023}, these seemingly generalising effects have been shown \citep{vonHausegger2024} to already be fully included in the \citet{Ellis1984} amplitude (Equation~\ref{eq:dipole_amplitude}) used all along. Specifically, as the catalogues treated here are redshift-integrated, so are these aforementioned redshift evolution effects, reducing any redshift evolving parameter (such as $x$ or $\alpha$) to a well-defined average, so long as this is measured at the corresponding flux limit of each catalogue. Central to this conclusion is the understanding that any luminosity function describing the sources in redshift must also describe the unique redshift-integrated source counts that are actually observed, which had not been considered in the proposition by \citet{Dalang2022} and \citet{Guandalin2023}. While additional, kinematic terms are expected in, e.g., redshift tomographic measurements \citep{vonHausegger2024a}, again the fully redshift-integrated catalogues as those here (barring the $z<0.02$ cut, which we estimate to have a negligible effect) are not known to require any corrections to Equation~\ref{eq:dipole_amplitude}, justifying our first defining assumption. 

Our second assumption states that the residual dipole is present in all surveys with the same amplitude and direction. Of course there is no way to know whether this assumption is accurate in light of the fact that we do not know what or if the residual dipole truly is. As mentioned above, this was motivated by the observation in \citet{Secrest2022}. In this sense, it is not surprising that we recover posteriors for $\beta$ that appear peaked close to the canonical velocity. Assuming the residual dipole to be different per survey, for instance by considering it to originate from random shot noise that, for the given catalogue sizes is not insubstantial, leads to our considered hypothesis (0). This hypothesis yields a recovered value for $\beta$ which lies at around twice the canonical value. While the posteriors and Bayes factors comparing the models with and without a residual dipole favour the former option, the evidence is not strong enough to be conclusive. Also, a residual dipole that somehow varies per catalogue, for instance as a function of frequency \citep[as suggested by][]{Siewert2020} invalidates this assumption, although this would of course still require that a non-kinematic component is present. Clearly, the natural degeneracy of $\mathcal{D}_{kin}$ and $\mathcal{D}_{resid}$ is only broken if an additional assumption on the stability of $\mathcal{D}_{resid}$ is made. However, should future theoretical investigations predict a common, non-kinematic contribution to the matter dipoles across samples, then our work offers a prediction of just this. 

\subsection{The multi-wavelength dipole}

In this work, we have used catalogues gathered at different wavelengths to constrain the dipole and its components. With this choice of catalogues, we have implicitly combined different objects: Quasars at mid-IR frequencies and predominantly AGN at the radio frequencies. Though they are observationally distinct, these source samples are all part of the active galaxies class, and therefore have similar redshift distributions with a median of $z\sim1$ \citep[e.g.][]{Secrest2021, Tiwari2016}. Going forward however, when fainter sources, especially in the radio, will be included in these studies, SFGs will begin to dominate the radio samples. These will present a wholly different source population, for which perhaps a residual dipole might deviate to a significant degree \citep[for a recent example see][]{Wagenveld2024a}. One might thus attempt to interpret $\mathcal{D}_{resid}$ as a clustering dipole, making studies of galaxy bias paramount \citep[cf.][]{Peebles2022a} in light of the diversity of galaxy types inherent to the different samples.

Nonetheless, this difference in frequency consequently means that instrumental systematics do not offer a plausible explanation for these dipoles, not only due to the vastly different experimental details of NVSS, RACS-low and CatWISE, respectively, but also the range in frequencies by which the radio and mid-IR measurements differ. The residual dipole entertained here hence is not to-be-interpreted as a systematic. Rather, it is a reframing of the upheld cosmic dipole anomaly in a quest of understanding its cosmological origin: Independent of the decomposition of the observed dipoles into kinematic and residual contributions, the conclusion that $\mathcal{D}_{obs}$ is too large to be reconcilable with the cosmological principle remains.

\subsection{Theoretical context}

If the kinematic dipole is that expected from the kinematic interpretation of the CMB dipole, as is hinted at by our analysis, the corresponding residual dipole is too large to be explained within the linear perturbations expected in the standard cosmological model. In other words, the clustering contributions to the number count dipole in the high-redshift samples investigated here, as predicted within \textLambda-CDM, are smaller than the inferred residual dipole seen by a factor of a few tens \citep[depending on the galaxy bias $b(z)$,][]{Gibelyou2012,Secrest2021}.

Thus, the question as to which phenomenon would produce a non-kinematic $\mathcal{D}_{resid}$ still stands. To understand the difference between CMB dipole and the corresponding expected dipole in galaxy number counts, one may consider the presence of large horizon-sized density perturbations. It is known that an adiabatic mode of such scale does not alter the CMB dipole, compared with what is expected due to our currently understood observer velocity \citep[e.g.][]{Grishchuk1978,Erickcek2008}. Instead, this sparked studies into iso-curvature modes \citep[e.g.][]{Gunn1988,Turner1991,Turner1992,Langlois1996,Erickcek2009} which do cause a sought-for difference, by adding a non-vanishing, intrinsic dipole component to that of the CMB. Regarding the number count dipole, however, \citet{Domenech2022} find that neither adiabatic, nor iso-curvature perturbations can be associated with a number count dipole as large as those observed, if $\beta$ remains at its canonical value. One must therefore, if considering a substantial non-kinematic $\mathcal{D}_{resid}$, speculate about other phenomenology at large scales, such as non-Gaussianity or effects of non-linear growth. 

\subsection{Outlook}

While awaiting larger cosmological surveys such as \textit{Rubin}-LSST, \textit{Euclid}, or SKA, we are currently in a situation where analyses of data sets in synergy offer the best attempt at understanding the origin of the cosmic dipole anomaly. As such, it is worth noting here that the excess dipole is not the only potentially problematic anisotropy that has been observed so far. Particularly noteworthy in this context are the anisotropies seen in the scaling relations of X-ray clusters \citep{Migkas2018,Migkas2021}, which are either linked to anisotropic Hubble expansion or bulk flows beyond the scale of what is expected in a homogeneous and isotropic Universe. Excess bulk flows have also been seen in other data sets \citep{Watkins2023}. It is very possible that these anisotropies are connected to the ones mentioned here as well as other observed cosmological inconsistencies \citep[see][for an overview]{Aluri2023}. However, considering the results obtained here, especially the constraints on $\beta$ obtained with hypothesis (i), a measured $\beta$ below the canonical velocity is fully consistent with our constraints, and could be obtained if the observed source population is also moving with respect to the CMB rest frame. This would be in line with an excess bulk flow, although it would have to be sustained to much larger scales than the currently available velocity data. 

In this work we considered combinations of three data sets that to-date have delivered the most robust measurements on the number counts dipoles, which in total add up to about 2.3 million sources. Yet, even having combined these current data sets, the width of the posteriors remained too broad to conclusively distinguish a kinematic from a non-kinematic origin of the observed matter dipoles. Nevertheless, we demonstrated the feasibility of this distinction exclusively with galaxy number counts by leveraging the sample-dependent kinematic matter dipole expectation, Equation~\ref{eq:dipole_amplitude}, and by considering different assumptions on an additional, non-kinematic, residual dipole component. 

\section{Conclusion}
\label{sec:conclusion}

In this paper we have leveraged the differences between the expected kinematic dipole amplitudes of different catalogues to try to isolate the kinematic dipole from a possible residual dipole effect. To this effect, we assumed that the only kinematic contribution to the number count dipole is given by the \citet{Ellis1984} amplitude (Equation~\ref{eq:dipole_amplitude}), which allowed us to separate kinematic from non-kinematic dipole contributions. Differences between the expectations for different catalogues are caused by different values of $x$ and $\alpha$ in each catalogue. To this end, we used the radio catalogues of NVSS and RACS-low, and the infrared CatWISE catalogue. These catalogues have previously yielded robust measurements of the cosmic number count dipole individually, and therefore formed our primary selection of data sets, whereas other data sets individually delivered less significant dipole measurements. Furthermore, with some minor cuts, the catalogues cover the whole sky twice over while being entirely independent. With this, the different catalogues could be combined for a joint dipole estimation as previously defined in \citet{Wagenveld2023b}.

Using the different expected kinematic dipole amplitudes for the catalogues, we separately fit for the kinematic and residual dipoles. Although freeing up both amplitudes and directions of these dipoles did not yield a converging estimate, letting the different dipole components have a common direction but different amplitudes did.  The resulting posterior indicated a preference towards a lower kinematic dipole component with $\beta < 2.35\times10^{-3}$ (95\% CL), consistent with the canonical velocity taken from the CMB dipole. Consequently, as per Equation~\ref{eq:dipole_components}, the remaining excess dipole amplitude was found to be in the form of a residual dipole, the origin of which is not known. Freeing up the parameters of the residual dipole while fixing those of the kinematic dipole to the CMB values, we found a residual dipole offset from the CMB direction by $39\pm8$ degrees, with an amplitude of $\mathcal{D}_{resid}=(0.81\pm0.14)\times10^{-2}$, deviating from a non-existent residual dipole by $5.5\sigma$. Evidently, this significance simply reflects a rephrasing of the notorious dipole anomaly. Furthermore, this interpretation appeared favoured compared to the other hypotheses according to the computed Bayesian evidences.

It is enticing to have found a preference towards the interpretation that the observed number count dipole in different radio-, infrared- and optical surveys consists of a kinematic component that matches the velocity derived from the CMB dipole along with a residual, non-kinematic dipole. The catalogues employed here do not have the sensitivity to fully break the degeneracy between the different dipole components, calling for a repeat measurement with a future set of deeper, but still independent catalogues. Deeper radio catalogues, both in the northern and southern hemispheres, would be particularly useful, as here CatWISE dominates the estimates, especially in terms of direction.   

\begin{acknowledgements}
We thank the anonymous referee for their insightful comments and feedback.
Some of the results in this paper have been derived using the \textsc{healpy} \citep{Zonca2019} and \textsc{HEALPix} \citep{Gorski2005} packages.
This research has made use of \textsc{topcat} \citep{Taylor2005}, \textsc{bilby} \citep{Ashton2019}, \textsc{emcee} \citep{Foreman-Mackey2013}, and \textsc{harmonic} \citep{McEwen2021}.
\end{acknowledgements}

\bibliographystyle{aa}
\bibliography{main_paper}

\begin{appendix}
\onecolumn
\section{Additional combined dipole estimates}
\label{app:additional_results}

\begin{figure*}[h]
    \centering
    \includegraphics[width=0.75\textwidth]{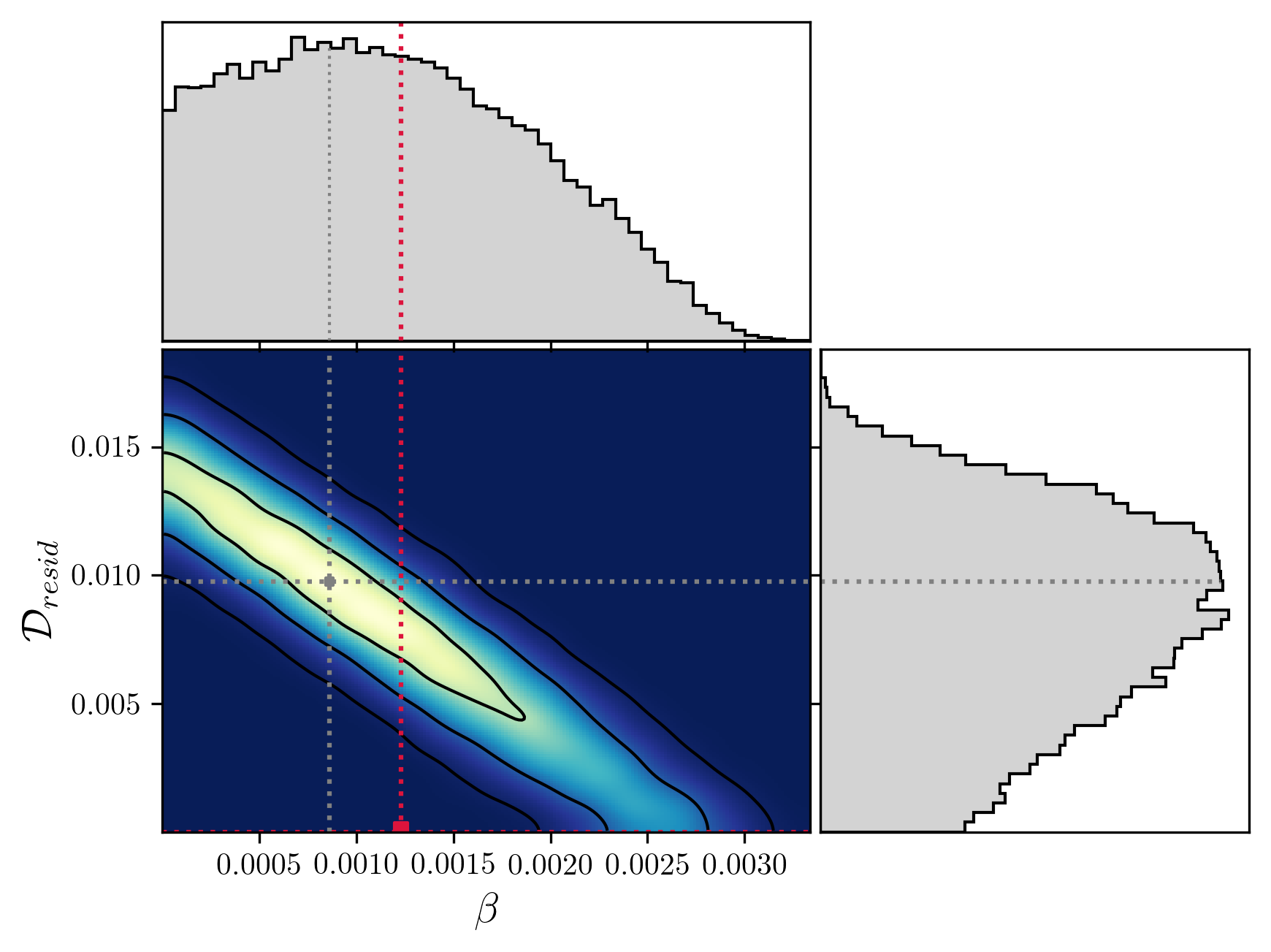}
    \includegraphics[width=0.75\textwidth]{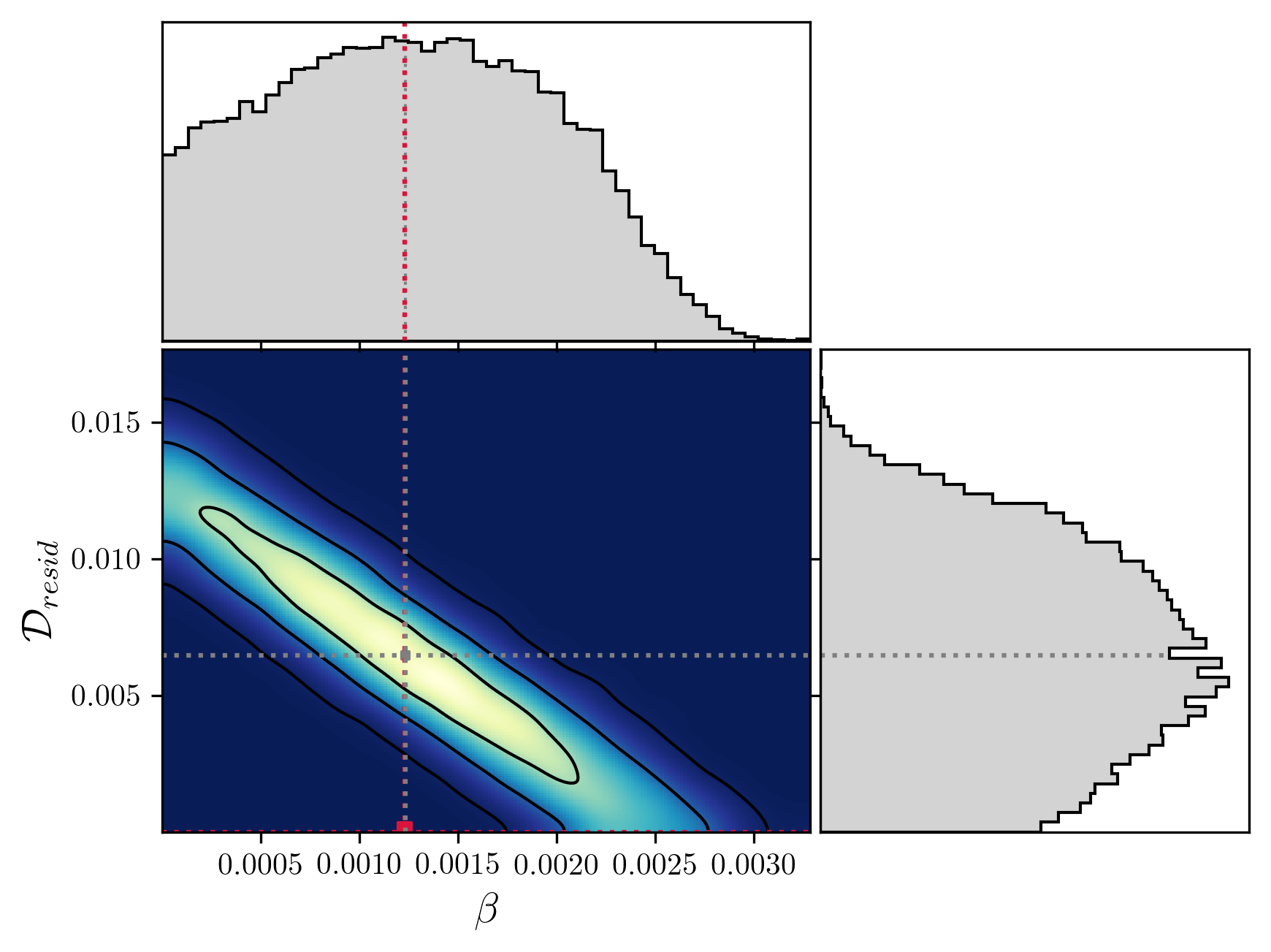}
    \caption{Posterior distributions of $\beta$ and $\mathcal{D}_{resid}$ from the combined estimate using NVSS and CatWISE (upper panel), and RACS-low and CatWISE (lower panel). The 1-,2-, and 3-$\sigma$ uncertainties are indicated by the black contours. The dotted line indicates the maximum posterior values for these parameters. The canonical CMB velocity of $\beta=1.23\times10^{-3}$ is indicated by the dotted red line, the red dot indicating the kinematic dipole expected in the standard cosmology, assuming a negligible structure dipole ($\mathcal{D}_{resid}\approx0$)}
    \label{fig:beta_dipole_app1}
\end{figure*}

\begin{figure*}[h]
    \centering
    \includegraphics[width=0.75\textwidth]{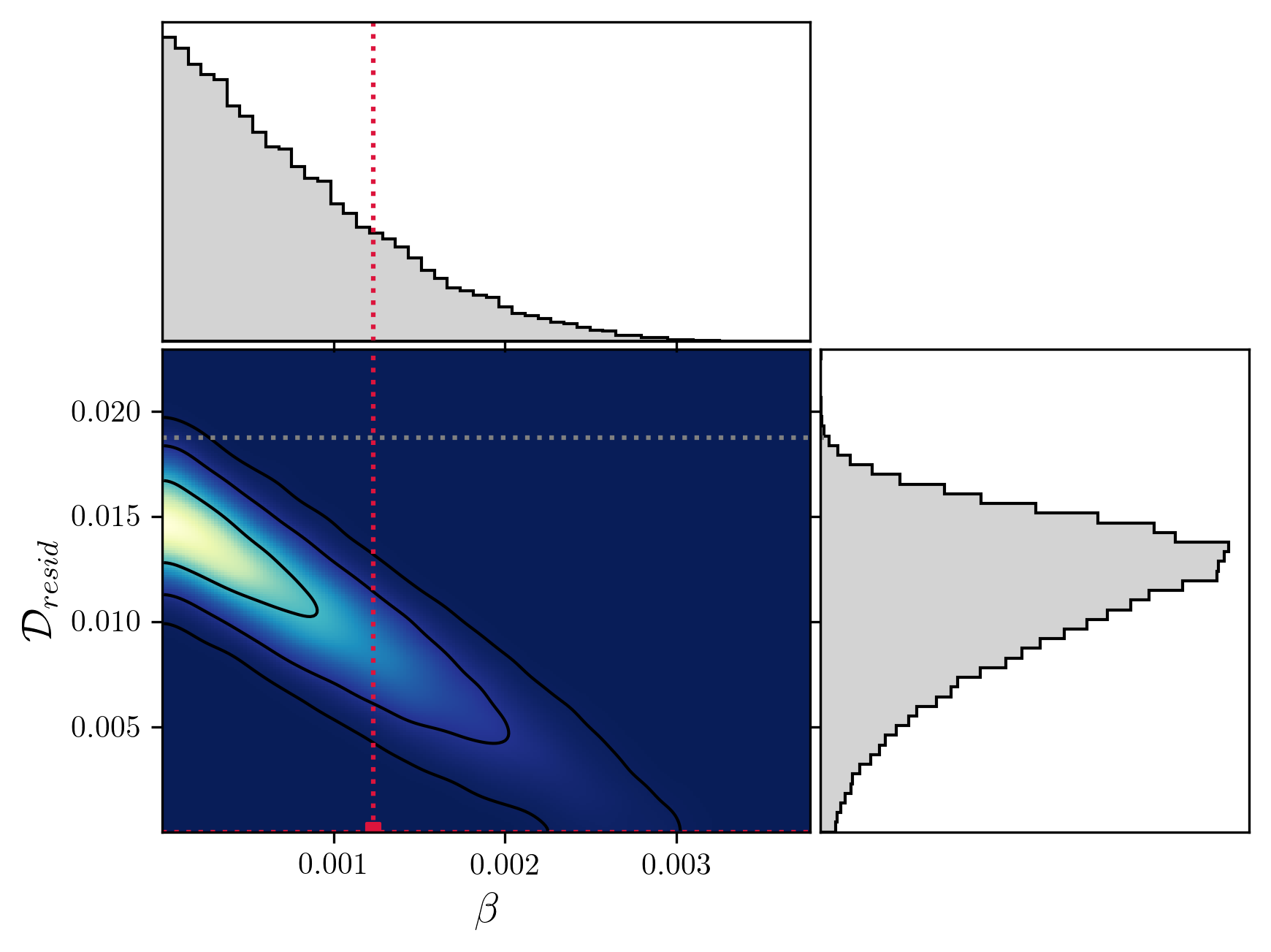}
    \includegraphics[width=0.75\textwidth]{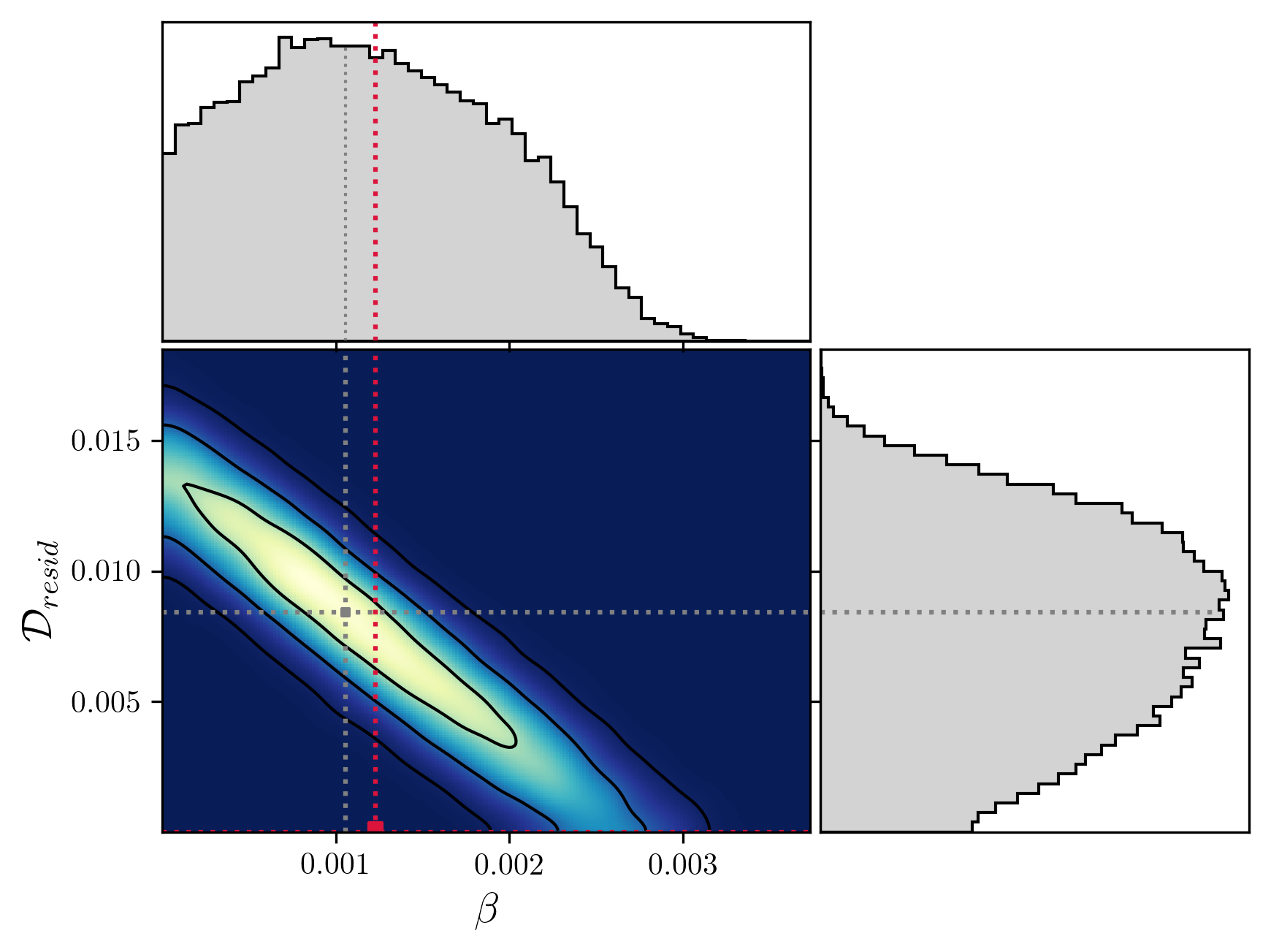}
    \caption{Posterior distributions of $\beta$ and $\mathcal{D}_{resid}$ from the combined estimate using NVSS, RACS-low, and CatWISE. In the upper panel, higher flux density cuts are applied, with 20 mJy for NVSS and RACS-low, and 0.09 mJy (W1 < 16.4) for CatWISE. In the lower panel, sources with $z<0.1$ in 2MRS have been removed from all catalogues. The 1-,2-, and 3-$\sigma$ uncertainties are indicated by the black contours. The dotted line indicates the maximum posterior values for these parameters. The canonical CMB velocity of $\beta=1.23\times10^{-3}$ is indicated by the dotted red line, the red dot indicating the kinematic dipole expected in the standard cosmology, assuming a negligible structure dipole ($\mathcal{D}_{resid}\approx0$)}
    \label{fig:beta_dipole_app2}
\end{figure*}
\FloatBarrier
\begin{table*}
    \renewcommand*{\arraystretch}{1.4}
    \centering
    \caption{Best fit dipole estimates for different data combinations and cuts. Where catalogues are combined only hypothesis (i) is tested, with results shown in Figures~\ref{fig:beta_dipole_app1} and \ref{fig:beta_dipole_app2}.}
    \begin{tabular}{l l c c c c c}
    \hline \hline
    Catalogue(s) & $S_0$ & $N$ & $\mathcal{M}$ & $\mathcal{D}$ & R.A. & Dec.\\
     & (mJy) & & counts/pixel & ($\times10^{-2}$) & (deg) & (deg) \\ \hline
    NVSS & 20 & 272,977 & $7.87\pm0.02$ & $1.60\pm0.34$ & $150\pm13$ & $-25\pm14$ \\
    RACS-low & 20 & 346,092 & $11.14\pm0.02$ & $1.68\pm0.27$ & $191\pm10$ & $9\pm13$ \\
    CatWISE & 0.09 & 1,216,501 & $52.49\pm0.05$ & $1.56\pm0.19$ & $141\pm6$ & $-6\pm7$ \\ \hline
    NVSS$^{\mathrm{a}}$ & 15 & 349,439 & $10.02\pm0.02$ & $1.17\pm0.30$ & $149\pm14$ & $-9\pm17$ \\
    RACS-low$^{\mathrm{a}}$ & 15 & 438,085 & $14.07\pm0.02$ & $1.25\pm0.24$ & $194\pm12$ & $5\pm15$ \\
    \hline
     &  & & $\beta$ & $\mathcal{D}_{resid}$ & & \\
     & & & ($\times10^{-3}$) & ($\times10^{-2}$) & & \\ \hline
    NVSS\,CatWISE & 15\,0.078 & 1,884,811 & $<1.85$ & $0.98\pm0.50$ & $144\pm6$ & $-11\pm6$ \\
    RACS-low\,CatWISE & 15\,0.078 & 1,973,705 & $1.23^{+0.87}_{-1.03}$ & $0.62^{+0.57}_{-0.44}$ & $152\pm6$ & $-10\pm6$ \\
    NVSS\,RACS-low\,CatWISE & 20\,20\,0.09 & 1,579,687 & $<0.90$ & $1.47_{-0.44}^{+0.20}$ & $144\pm5$ & $-8\pm6$ \\
    NVSS\,RACS-low\,CatWISE$^{\mathrm{a}}$ & 15\,15\,0.078 & 2,032,858 &  $1.06\pm0.93$ & $0.84\pm0.50$ & $145\pm5$ & $-9\pm6$ \\ \hline
    \end{tabular}
    \tablefoot{$^{\mathrm{a}}$ Excluded sources matched to the 2MRS catalogue at $z < 0.1$.} 
    \label{tab:results_app}
\end{table*}

\end{appendix}

\end{document}